\documentstyle[aps,twocolumn,epsfig]{revtex}
\begin{document}
\draft
\preprint{\today}
\title {Isovector and Isoscalar superfluid phases in rotating nuclei\\}
\author{Javid A. Sheikh$^{(1,2)}$ and Ramon Wyss$^{(1)}$}
\address {$^{(1)}$ Royal Institute of Technology, Stockholm,
Sweden \\
$^{(2)}$Physik-Department, Technische Universitat Munchen,
85747, Garching, Germany\\
}

\maketitle

\begin{abstract}
The subtle interplay between the two nuclear superfluids, isovector T=1 and
isoscalar T=0 phases, are investigated in an exactly soluble model. 
It is shown that T=1 and T=0 pair-modes
decouple in the exact calculations with the T=1 pair-energy 
being independent of the T=0 
pair-strength and vice-versa. In the rotating-field, the isoscalar
correlations remain constant in contrast to the well known
quenching of isovector pairing.
An increase of the isoscalar (J=1, T=0) pair-field
results in a delay of the bandcrossing frequency. This
behaviour is shown to be present only near the N=Z line and
its experimental confirmation would imply a strong signature for
isoscalar pairing collectivity. 
The solutions of the exact model are
also discussed in the Hartree-Fock-Bogoliubov approximation.

\end{abstract}


There is  overwhelming evidence that the isovector, T=1 pairing
field among identical nucleons is an essential component of the 
nuclear mean-field potential. The bulk of  
nuclear ground-state properties, like the odd-even mass differences
and the moments of inertia of deformed nuclei can be accounted
for by considering nucleons to be in a superfluid
(T=1, J=0) paired-phase\cite{[BM75]}. 
These effects have been studied  mostly in 
heavier nuclei with N$>$Z, where the Fermi surfaces 
of protons and neutrons lie in different major
shells. 

In recent years, however, due to a substantial progress achieved in
the sensitivity of the detecting systems it has been possible to study
nuclei near the N$=$Z line in the mass A$=$70 and 80 regions.
Furthermore, with the
availability of radioactive beams these studies are expected to
reach even heavier N$=$Z nuclei. For these
nuclei, one expects the  pairing between
protons and neutrons to become  important, since the Fermi
surfaces of both protons and neutrons lie in the same major shell.

The role 
of the isovector T=1 pairing between protons and neutrons in the
low-spin regime has been discussed in recent studies
by \cite{[Rud96],[Fra99]}.
The importance of the isoscalar T=0 pairing can be inferred from
masses\cite{[Sat97]} and studies of
high-spin states\cite{[Mul81],[Ter98],[Sat00]}. 
However, most of
these studies are based on the mean-field approximation which
predict a transitional behaviour for rotating nuclei
for the T=1 and T=0 pair-fields as
a function of the rotational frequency and the strength of the T=0 
interaction\cite{[Sat97]}.

The purpose of the present study is to examine properties of 
the isoscalar and isovector correlations  within
an exactly soluble model of a deformed sinlge-j shell
and to compare to the predictions 
of the mean-field HFB approximation. 
The observable consequences of the T=0
pair-field which have remained illusive are also discussed
in the present study.

The model hamiltonian consists of a cranked deformed
one-body term and a scalar two-body interaction\cite{srn90,she90}
\begin{equation}
H^\prime = h^\prime + V_2,
\end{equation}
where,
\begin{equation}
h^\prime = h_{def} - \omega J_x,
\end{equation}
with
\begin{equation}
h_{def} = -4 \kappa { \sqrt { 4 \pi \over 5 }} \sum_{i j} 
        < j | Y_{20} | i>
         \delta_{\tau_i \tau_j} \delta_{m_i m_j} c^\dagger_{j}c_{i}.
\end{equation}
The labels $i,j,...$ denote the magnetic quantum-number ($m$) of the j-
shell and the isospin projection quantum-number  $\tau$
[$\tau$=1/2 (neutron) and -1/2(proton)]. 
The deformation energy $\kappa$ is equal to  the usual
deformation parameter $\beta$ by $\kappa = 0.16 \hbar \omega (N+3/2) \beta$ in units of G
(ref. \cite{swv98}). 
The two-body interaction in Eq. 1 is given by
\begin{equation}
V_{2}={\frac{1}{2}}\sum_{JMTT_z}E_{JT}^{{}}A_{JM;TT_z}^{\dagger }A_{JM;TT_z}^{{}},
\label{E503}
\end{equation}
with $A_{JM;TT_z}^{\dagger }=(c_{j { 1\over 2}}^{\dagger }
c_{j{1 \over 2}}^{\dagger })_{JM;TT_z}$ and $%
A_{JM;TT_z}^{{}}=(A_{JM;TT_z}^{\dagger })^{\dagger }$. For the
antisymmetric-normalized two-body matrix-element ($E_{JT}^{{}}$), we use
the delta-interaction which for a single j-shell is given by \cite{gb} 
\begin{equation}
E_{JT}^{{}}=-G{\frac{(2j+1)^{2}}{2(2L+1)}}\left\{ \left[ 
\begin{array}{ccc}
j~ & ~j~ & J \\ 
\frac{1}{2}~ & -\frac{1}{2} & 0
\end{array}
\right]^{2} + { 1 \over 2}\{1+(-1)^{T}\}
\left[
 \begin{array}{ccc} 
j~ & j~ & J \\ 
\frac{1}{2}~ & \frac{1}{2}~ & 1
\end{array}
\right]^{2}
\right\}
\label{E504}
\end{equation}
where the bracket $[~~]$ denotes the Clebsch-Gordon coefficient.

As mentioned in the introduction, one of the objectives of the present 
work is
to investigate the HFB approximation. In the following, we present some basic 
HFB formulae, for details see for instance ref.\cite{[RS80]}. 
The HFB equations are given by

\begin{equation}\label{HFB1}
{\cal H}'\left( \begin{array}{c} U\\V \end{array}\right)=E'_i \left( 
\begin{array}{c} U\\V \end{array}\right),
\end{equation}
where
\begin{eqnarray}
{\cal H}'=&
\left( \begin{array}{cc} h^\prime~~~ & \Delta_{} \\
-\Delta^\ast_{}~~~&-(h^\prime)^\ast 
\end{array}\right).\label{hfbh}&
\end{eqnarray}
with
\begin{eqnarray}
h^\prime_{ij} &=& \epsilon^\prime_{ij}+\Gamma_{ij},\\
\epsilon^\prime_{ij} &=& < i | h_{def}|j> - ( \lambda_p Z +
\lambda_n N +\omega m_i) \delta_{ij},\\
\Gamma_{ij} &=& \sum_{kl} <ik|v_a|jl> \rho_{lk},\\
\Delta_{ij} &=& {1 \over 2} \sum_{kl} < ij|v_a|kl> \kappa_{kl}.
\end{eqnarray}
\begin{equation}
\rho =V^{\ast }V^{T},\quad \quad \kappa =V^{\ast }U^{T}=-UV^{\dagger }.
\label{HFBN}
\end{equation}
In order to evaluate the angular-momentum dependence of the pair-energy, we
define the coupled pair-field through
\begin{eqnarray}
\Delta_{ij} = \sum_{JMTT_z} 
\left[
\begin{array} {ccc} j~~ & j~~ & J\\
                   m_i~~ & m_j~~ & M\\
\end{array}
\right]
\left[
\begin{array} {ccc} {1 \over2}~~ & {1 \over 2}~~ & T \\
                   \tau_i~~ & \tau_j~~ & T_z \\
\end{array}
\right]
\Delta_{JT},
\end{eqnarray}
with
\begin{eqnarray}
\Delta_{JT} = E_{JT} \sum_{ij}\\
\left[
\begin{array} {ccc} j~ & j~ & J\\
                   m_i~ & m_j~ & m_i+m_j\\
\end{array}
\right]
\left[
\begin{array} {ccc} {1 \over 2} & {1 \over 2} & T \\
                   \tau_i & \tau_j & \tau_i+\tau_j \\
\end{array}
\right]
\kappa_{i j}.
\end{eqnarray}
The pair-energy can now be expressed in terms of the coupled pair-fields as
\begin{equation}
E_{pair} = { 1 \over 2} \sum_{JT} { \Delta_{JT} \Delta_{JT}^{*} 
\over E_{JT} }.
\label{EEP}
\end{equation}
The above expression is quite useful since in the exact calculations there
is no gap parameter $\Delta$, but one may associate ``$E_{pair}$'' with the
expectation value of the two-body residual interaction, $V_2$, Eq.\ref{E503}
To obtain the $\Delta$-value from the exact analysis,
Eq.\ref{EEP} is then simply inverted.

The HFB solutions have been obtained by solving the 
Eqs. (\ref{HFB1}-\ref{HFBN})
self-consistently. In order to treat both the T=0 and the T=1 pair-fields
simultaneously, it is necessary to define complex HFB potentials, 
since the symmetries of the T=1 and
T=0 n-p fields are different\cite{[Che67]}.
The initial complex HFB wavefunctions have been constructed by using
the expressions for real and imaginary $V$'s and $U$'s of the 
HFB transformation
in terms of the pair-gaps\cite{[Che67]}.
We would like to mention that no symmetry restrictions have been imposed 
on the HFB wavefunction since it is known that symmetries lead to exclusion
of particular correlations. For more details concerning the HFB-transformation
in the presence of both T=1 and T=0 pairing, we refere the reader 
to refs.~\cite{[Goo79],[Goo99]}.

\begin{figure}[htb]
\noindent
\hspace*{-1.0cm}
\vspace*{-3.0cm}
\epsfig{file=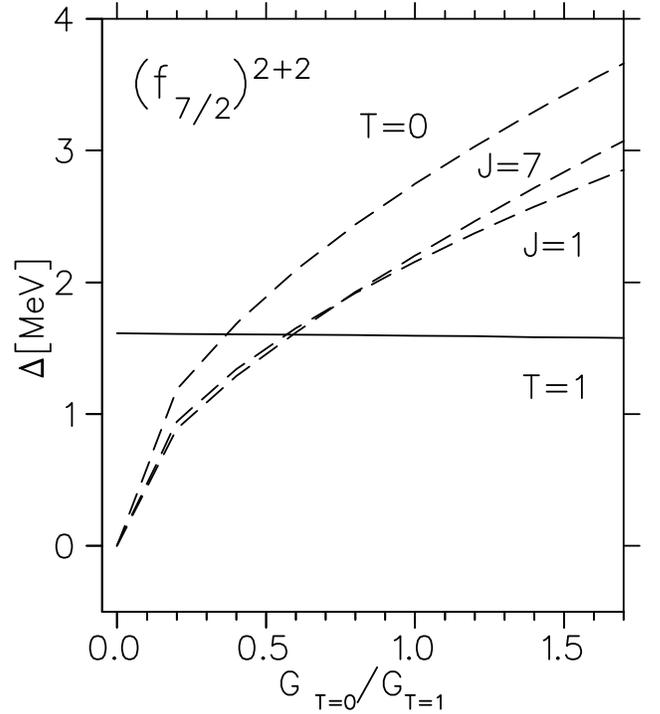,scale=0.57}
\caption{The exact
single-j shell model pairing-gaps as a function of the
T=0 strength for a system with 2-protons and 2-neutrons in 
$f_{7/2}$ shell.
}
\label{figure.1}
\end{figure}

Several mean-field studies show 
that the T=0 and T=1 pairing-modes are exclusive in the 
BCS-approximation\cite{[Sat97],[Wol71]}. 
The system is always choosing the mode that generates the lowest energy, 
which in the case of equal weight for each pair results in either  
T=0 and T=1 pairing\cite{[Sat00]}.
In the presence of approximate particle-number projection,
the two modes coexist, but only above  a critical 
strength\cite{[Sat00],[Sat97]}. 
Using a more complex model space also results in the possibility
of mixed solutions\cite{[Ter98],[Goo99]}.
The question, 
therefore, arises whether the exclusiveness is persistent in an exact model.
Fig.~1 
shows the size of the T=1 
correlations as a function of increasing T=0 strength in the exact
analysis. The figure clearly shows 
that the two modes are essentially independent. 
There is no
critical strength for either pairing mode
and therefore one expects to have both modes present in
nuclei.  
From this we can conclude that the exclusion between the two modes is 
a mean-field effect. It also implies that atomic nuclei exhibit the unique
possibility of exhibiting two different pairing condensates  
simultanously.

\begin{figure}
\noindent
\hspace*{-0.5cm}
\vspace*{-2.cm}
\epsfig{file=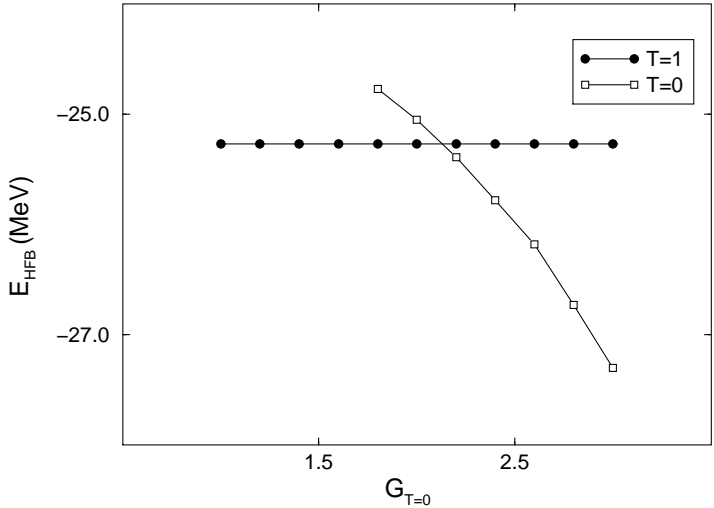,scale=0.55}
\caption{The  HFB pairing energy for 4-protons and 4-neutrons
as a function of the T=0 strength. 
}
\label{figure.2}
\end{figure}

In order to explore further the mutual exclusiveness of the T=1 and T=0
pair-fields obtained in earlier studies, we have studied the 
HFB solution as a function of the strength of the T=0 interaction.
The results are presented in Fig. 2.  For the
normal strength $G_{T=0}=1$, the solution corresponds to a T=1 pair-field.
With increasing $G_{T=0}$ the HFB energy remains constant which is
obvious since the solution has only the T=1 component and there is no
T=0 component. The T=0 solution shown in Fig. 2 has been obtained by 
solving the HFB equations for a very large value of $G_{T=0}$ 
($G_{T=0}=2.8$)
and then using
this solution for lower values of $G_{T=0}$. 
In this manner, it was possible to obtain a T=0 solution also below the 
critical point, see Fig.~2.
We note from Fig. 2 that the two solutions
coexist for most of the $G_{T=0}$ values. 
They represent two different 
solutions of the HFB equations. 

The exact solution, presented in Figs. 1
contains both the T=0 and T=1 pair-modes, whereas HFB gives two
 separate solutions, 
corresponding to either T=0 or T=1 pair-fields.
The difference between the two models
resides in the fact that in the exact model, the two-body 
interaction always is a scalar whereas in the HFB-aproximation,
the pairing potential is either a T=0 or T=1 field, with the
corresponding symmetry. 
Our analysis shows that starting from a certain solution, 
with a given symmetry, this symmetry propagates to the next 
solution (with different $G_{T=0}$), analoguous to other self-consistent 
symmetries of the HFB hamiltonian, see e.g. the discussion in \cite{[Goo79]}.
The different pair-fields appear as independent of each other. Our 
results further indicate, that for a certain strength of the  $G_{T=0}$ pair 
field, energy can be gained. This conforms with earlier results to associate 
the Wigner energy with T=0 pair correlations\cite{[Sat97]}.

\begin{figure}[htb]
\noindent
\hspace*{-1.5cm}
\epsfig{file=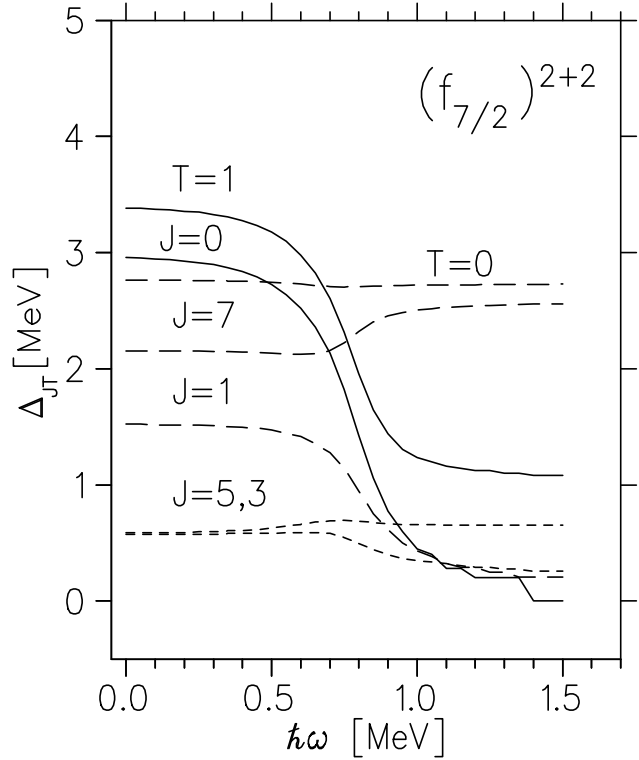,scale=0.85}
\caption{Behaviour of the exact shell model pair-gaps as a 
function of rotational frequency
$\hbar\omega$ for 2+2 particles
in the $f_{7/2}$ shell. The solid (dashed) lines represent the T=1 (T=0)
part of the pairing-energy. For the case of T=0, we show all individual
components of the force, clearly demonstrating the importance of
the different $J$'s. In contrast,
the T=1 force is dominated by the J=0 component.
}
\label{figure.3}
\end{figure}

As a next step, we consider the response of the nuclear 
pair-potential to the rotating fields. 
In Fig. 3, we show the total pair-field (Eq. 15)
as well as selected individual $(J,~T)$ contribution as a function of the
rotational frequency ( $\hbar\omega$ ) for 4 particles 
(2 protons and 2 neutrons)
in the $f_{7/2}$ shell. First of all, we may note the
distinct difference between the
T=1 and T=0 pairing fields. Whereas the T=1 field is dominated by one 
component with J=0, the T=0 mode is dominated by the J=1 and 
$J=2j$ part of the interaction, also the intermediate spins $J=3,~5$ 
play a role.
This already  indicates that a discussion of a pairing force restricted 
to $L=0$ may be appropriate for the T=1 part of the interaction, but not for
T=0, see also ref.~\cite{[Sat97a],[Sch76]}.

As we increase the rotational frequency,  
the T=1 pairing-correlations
(solid line) reveal the well known drop due to particle-alignment
from the $f_{7/2}$-shell at around $\hbar \omega = 0.7G$. At this crossing
point, the yrast band changes character from the paired $(J=0)$ configuration
to the aligned $(J=M_x=6+6)$ state.

Similar calculations were performed 
also for the case of the (4+2) and (4+4) systems.
Qualitatively, they all show the same trend, where of course the 
size of the drop in correlation energy depends on the number of particles
present in the single-$j$ shell. 
For  (4+2) system, the correlations of the J=0
component only for one-pair disappear whereas the drop for the
 4+4 particles  is less pronounced. This is
due to the fact that only one-proton and one-neutron pair have aligned 
at the first crossing. Hence, the J=0 correlations are still
active for the remaining two-pairs. For higher frequencies,
the next pair will align, and then the J=0 (and in consequence)
the T=1 correlations will drop in a similar fashion as for the
system with one-proton and one-neutron pair only. 
The important message remains,
as is evident from Fig. 1, that the T=1 field is largely
built up from the J=0 pair-correlations, that are diminished
in the process of  particle alignment. Although, the components
with higher-$J$ contribute at higher values of angular-momentum,
the T=1 correlations are strongly reduced by
the rotational motion. 

In contrast, the T=0 correlations evolve quite differently with rotational 
frequency.
The contribution of the coupling to low-$J$, like the J=1 pairs, behave
similar to the coupling to J=0. This is quite natural, since they are built 
up by pairs of $L=0$ and $L=2$.
However, although the contribution of
the J=1 to the T=0 correlations drop in a similar fashion
as the J=0, the value of the total T=0 correlations remain
essentially  unchanged. Apparently, the part
that is lost by J=1 and $J=3$  is gained by $J=7$ and
$J=5$. This implies, that the high-$J$ components of the T=0
correlations compensate the loss of  the low-$J$.
This feature appears to be independent of
the number of particles in the system. It means that for
a given interaction in the pp-channel, the total T=0 correlations
remain almost unaffected by rotation. 
The presence of increasing $L$-values in the pairing field will
affect deformation properties. This is what one expects in a
fully self-consistent approach, which of course is
beyond our present model analysis. Note that a recent analysis within the
Monte Carlo Shell model shows that at high angular momenta, the T=0 
correlations with $2j$ increase\cite{[Dea97]}.

From the above analysis, one may conclude that the T=0 correlations are not
able to affect rotational properties, since the increase in the stretched 
$J=2j$ component is exactly nullified by the decrease of the J=1 
part, see also the discussion in Ref.~\cite{[Fra99a]}.
Indeed, these are the results e.g.
for the  $f_{7/2}$ shell
where one is dealing with a ``single-j'' shell.
However, in heavier nuclei, when $Z>28$, the active 
shell is composed of, e.g., $p{_3/2},~f_{5/2}, ~p_{1/2}$ and $g_{9/2}$.
For those cases, the J=1 part of the T=0 interaction becomes more 
coherent, since every subshell can contribute. In contrast,  the $J=2j$ 
components become fragmented, since they have a different
value  for each subshell. Therefore, one may expect a different response
of the T=0 pair field to rotation in heavier nuclei. 
Since we are dealing in our model with a single-j
shell it is not possible to deal with such a case. One may, however,
mimick in an adhoc way the region beyond $Z=28$ by
increasing the strength of the J=1 part of the interaction.

\begin{figure}[htb]
\hspace*{-1.7cm}
\vspace*{-5.cm}
\epsfig{file=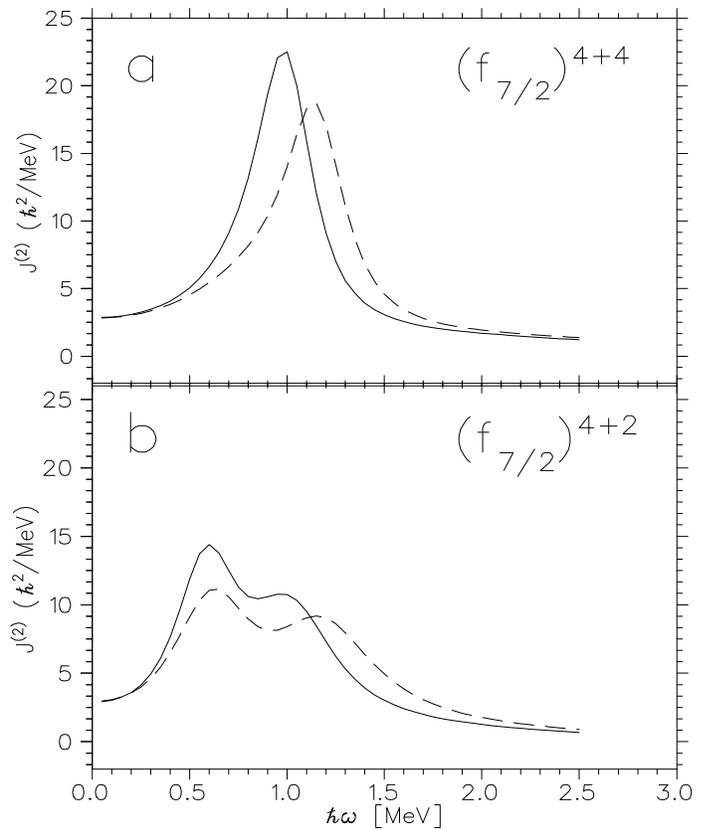,scale=0.68}
\vspace*{4.cm}
\caption{The dynamical momentof inertia, $J^2=dI/d\omega$, as a function of
frequency. Solid line correspond to standard single-j shell calculations,
whereas the dashed line depicts the case where the J=1 part of the 
interaction is increased by a factor of 2. Note the difference between
the 4+4 and the 4+2 system} 
\label{figure.4}
\end{figure}

The effect of a redistributed strength of the T=0 correlations, where the
J=1 part has been increased by a factor of two, is shown in Fig.~5. Indeed,
the crossing frequency is shifted. In other words, a coherence of J=1,
T=0 pairs results in a change of the crossing frequency.
What is even more striking is that this effect is suppressed when $N\neq Z$.
In Fig.~5, we show the case of (2+4) nucleons in the $f_{7/2}$ shell and, 
indeed, the first crossing frequency remains essentially unchanged.
This feature persists also in the HFB approximation.
Although our model is highly simplistic, one can certainly conclude that
T=0, J=1 collectivity results in a shift of the crossing frequency to 
higher values and that this property is expected to be present also
in more 
realistic calculations. Of course, as discussed above, there are other
factors that affect the crossing frequency, like 
the deformation which in turn can be 
influenced by the T=0 pairing field. 

A shift of the crossing frequency
has been reported for the case of the N=Z nucleus 
$^{72}$Kr\cite{[Ang97]}. 
There have been efforts to explain this shift
of the crossing in terms of T=1 np-pairing. 
Since to a very good approximation,  the 
nuclear force is charge independent, 
only the total isospin $T$ matters for the interaction, not the
projection of isospin ($T_z$). This is analoguous
to the assumption that the nuclear force
does not depend on the angular-momentum projection 
$J_z$, only on total $J$. This basic assumption implies 
that the T=1 pair-gaps are not 
affected by rotation in isospace, i.e. the total $T=1$ pair gap 
($\Delta_{nn}^2 + \Delta_{pp}^2 +\Delta_{np}^2$) 
is an invariant quantity\cite{[Sat97],[Fra99a]}. 
  In an attempt of Ref.~\cite{[Kan98]}
to account for the shift of the crossing frequency,
the T=1 $\Delta_{np}$ pair-gap was simply increased 
from 0 to a value of 2.5~MeV. Such an increase strongly violates
charge independence. Following the arguments given above,
one could as well increase the nn- or pp-pairing gaps.
Of course, any increase of the
T=1 pairing energy will result in a shift of the crossing frequency
but this has nothing to do with np-pairing.

In summary, we have studied the competition between the T=0 and T=1
pair-fields in an exactly soluble deformed single-j shell model. It
is shown that the HFB approach gives rise to two decoupled solutions
corresponding to T=1 and T=0 modes. Although, in the exact shell model
analysis, the solution contains both T=0 and T=1 modes, the two
modes are independent with T=1 pair-energy independent of the strength
of the T=0 correlations and vice-versa. 
The T=0 correlations in a single-j shell have a complex 
structure where the total amount is not affected by rotation. 
For  realistic 
cases in heavy nuclei (Z$>$28), with several j-shells, the J=1 part will 
effectively acquire a larger 
strength. It has been demonstrated that increasing the value of the (T=0, J=1)
pair-strength results in a shift of the bandcrossing frequency. 
Such a shift of the crossing frequency in heavy N=Z nuclei, therefore,
is an indication of the collective (T=0, J=1) correlations. 

This work has been supported by the G{\"o}ran Gustafsson Foundation and
the Swedish Natural Research Council (NFR).

\baselineskip = 14pt
\bibliography{pn}
\bibliographystyle{unsrt}

\end{document}